\documentclass[letterpaper, 10 pt, conference]{ieeeconf}  %
\IEEEoverridecommandlockouts                              %

\usepackage{makecell}

\usepackage{hyperref}
\usepackage{graphics} %
\usepackage{graphicx}
\usepackage{epsfig} %
\usepackage{amsmath} %
\usepackage{amssymb}  %
\usepackage{lipsum}
\usepackage{cleveref}
\usepackage{nicefrac}
\usepackage[table]{xcolor}
\usepackage{bm}
\usepackage{algorithm}
\usepackage{siunitx}
\usepackage{cite}
\usepackage[noend]{algpseudocode}

\newtheorem{definition}{Definition}%
\newtheorem{theorem}{Theorem}%
\newtheorem{proposition}{Proposition}%

\newtheorem{remark}{Remark}
\newcommand\real[1]{\mathbb{R}^{#1}}
\newcommand{\tf}[1]{\mathbf{#1}}
\newcommand{\ttf}[1]{\boldsymbol{#1}}
\newcommand{\spreg}{\operatorname{SpRegret}}
\newcommand{\sparse}{\operatorname{Sparse}}
\newcommand{\struct}{\operatorname{Struct}}
\DeclareMathOperator*{\st}{s.t.}

\DeclareMathOperator*{\H2}{\mathcal{H}_2}
\DeclareMathOperator*{\Hinfty}{\mathcal{H}_\infty}
\newcommand{\norm}[1]{\left\lVert#1\right\rVert}

\newcommand{\binmat}[1]{\mathbb{B}^{#1}}
\usepackage{nicematrix}
\usepackage{tikz}
\newcommand\scalemath[2]{\scalebox{#1}{\mbox{\ensuremath{\displaystyle #2}}}}
\definecolor{lightgray}{gray}{0.9}
\definecolor{darkgreen}{rgb}{0.0, 0.4, 0.0}

\newcommand{\preprintswitch}[2]{#2} %

\title{\LARGE \bf   
Closing the Gap to Quadratic Invariance: a Regret Minimization Approach to Optimal Distributed Control
}

\author{Daniele Martinelli, Andrea Martin, Giancarlo Ferrari-Trecate, and Luca Furieri%
\thanks{D. Martinelli, A. Martin, G. Ferrari-Trecate, and L. Furieri are with the Institute of Mechanical Engineering, EPFL, Switzerland. E-mail addresses: \{daniele.martinelli, andrea.martin, giancarlo.ferraritrecate, luca.furieri\}@epfl.ch.}
\thanks{ Daniele Martinelli and Luca Furieri are grateful to the Swiss National Science Foundation (SNSF) for the Ambizione grant PZ00P2\textunderscore208951. This research is also supported by the SNSF under the NCCR Automation (grant agreement 51NF40\textunderscore 80545).}
}
\begin{document}
\maketitle
\thispagestyle{empty}
\pagestyle{empty}
\begin{abstract}

In this work, we focus on the design of optimal controllers that must comply with an information structure. State-of-the-art approaches do so based on the $\H2$ or $\Hinfty$ norm to minimize the expected or worst-case cost in the presence of stochastic or adversarial disturbances. 
Large-scale systems often experience a combination of stochastic and deterministic disruptions (e.g., sensor failures, environmental fluctuations) that spread across the system and are difficult to model precisely, leading to sub-optimal closed-loop behaviors.
Hence, we propose improving performance for these scenarios by minimizing the regret with respect to an ideal policy that complies with less stringent sensor-information constraints. This endows our controller with the ability to approach the improved behavior of a more informed policy, which would detect and counteract heterogeneous and localized disturbances more promptly. Specifically, we derive convex relaxations of the resulting regret minimization problem that are compatible with any desired controller sparsity, while we reveal a renewed role of the Quadratic Invariance (QI) condition in designing informative benchmarks to measure regret.
Last, we validate our proposed method through numerical simulations on controlling a multi-agent distributed system, comparing its performance with traditional $\H2$ and  $\Hinfty$ policies.
\end{abstract}

\section{Introduction}
Control of large-scale systems, such as
smart grids~\cite{molzahn2017survey} or traffic systems~\cite{zheng2016distributed}, requires communication among multiple interacting agents to ensure efficient and safe operation.
A significant challenge arises from the incomplete information available to each agent regarding the overall system state. This partial communication can be due to various factors, including privacy concerns, geographic dispersion, and the inherent difficulties of establishing a reliable communication network.
 Designing optimal control policies complying with specific information structures is a well-known challenge, even in seemingly straightforward scenarios, as highlighted by the classic work~\cite{witsenhausen1968counterexample}.

The design phase concerning large-scale systems primarily revolves around addressing two fundamental challenges: first, how to parameterize controllers to comply with a given information sparsity, and second, how to formulate a metric that accounts for disturbances and the effect of their propagation throughout the dynamics of the distributed system.

Regarding the first challenge, for linear dynamical systems, a landmark contribution was given in~\cite{rotkowitz2005characterization}, where the authors proposed the notion of Quadratic Invariance (QI), a sufficient~\cite{rotkowitz2005characterization} and necessary~\cite{lessard2011quadratic} condition for enabling an exact convex reformulation of sparse linear controller synthesis. However, the QI condition can be overly restrictive, i.e., not fulfilled by several systems with highly intertwined dynamics.
To address this issue, \cite{furieri2019unified} introduced communication channels between controllers to effectively restore the QI condition.
The works \cite{wang2019system,furieri2019separable,furieri2020sparsity} presented convex optimization methods for designing sparse closed-loop dynamics and sparse controllers, even when the QI condition does not hold.
Within optimal distributed control, \cite{arastoo2016closed} addressed how to promote sparsity in closed-loop controllers while minimizing performance degradation compared to a centralized policy. Building on this, works \cite{jensen2021explicit} and \cite{fattahi2018transformation} proposed methods for the explicit parameterization of sparse $\H2$ controllers. \cite{jensen2021explicit} focused on physically decoupled systems, while \cite{fattahi2018transformation} derived control policies that closely resemble an optimal centralized one under specific conditions.

Concerning the second challenge of choosing an appropriate performance metric, traditional control techniques, such as $\H2$ and $\Hinfty$, often rely on certain assumptions about the nature of disturbances to achieve optimal control policies \cite{hassibi1999indefinite}. $\H2$ treats disturbances as stochastic noise, while $\Hinfty$ considers them as adversarial attacks.
However, these assumptions can be unrealistic and conservative in large-scale scenarios, where disturbances are 1) difficult to model due to the complex dynamics of small local mismatches propagating at large, and 2) difficult to localize, as they may affect agents at unpredictable locations and times.
This requires the development of novel control strategies capable of accommodating such non-standard disturbances.

\textbf{\emph{Contributions}}.
To improve the ability of the optimal policy to respond quickly to disturbances with unknown locations and nature, our idea is to close the gap to the performance of an \emph{oracle}, that is, a benchmark policy that possesses more sensor measurements.
By minimizing the worst-case difference in cost with this oracle, our controller design encourages emulation of its behavior, potentially leading to improved performance.
This new metric is inspired by the recent works on regret technique~\preprintswitch{\cite{sabag2021regret,goel2023regret,martin2022safe}}{\cite{sabag2021regret,goel2023regret,martin2022safe,martin2023regret}}, which are however limited to centralized control scenarios and focus on a temporal notion of regret.

In this work, we first analyze the conditions for the oracle to express an improved performance to be emulated. We term a spatial regret metric satisfying these capabilities as \textit{well-posed}. This analysis is achieved by revealing a renewed role of the QI condition within the well-posedness of the proposed metric.
Finally, we provide a convex reformulation for designing regret-optimal controllers with arbitrary sparsity structures, optimizing them to close the performance gap to an ideal QI subspace that encodes richer information for the control policy.
To illustrate the real-world relevance of our approach, we provide numerical examples involving a multi-agent scenario of a multi-mass spring-damper system that requires to be controlled in the presence of non-standard disturbances.

\textbf{\emph{Structure}}. 
The paper is organized as follows.
In \Cref{sec:problem_formulation}, we define the problem and introduce the novel spatial regret metric.
In \Cref{sec:main_results}, we prove the well-posedness of the spatial regret.
Moreover, we formulate the spatial regret optimization problem in a convex form.
Lastly, in~\Cref{sec:numerical_results} we report numerical examples to show the performance of spatial regret in the control of a distributed system.

\textbf{\emph{Notation}}. 
$\real{a \times b}$ represents the set of real matrices with dimension $a \times b$.
$\mathbb{N}^+$ denotes the set of non-zero natural numbers.
For a matrix $Y \in \real{a \times b}$, $Y_{i,j}$ indicates its element in row $i$ and column $j$. 
Given the vector $x\in\real{a}$, $x^i$ is used to indicate the $i^{th}$ component of $x$, with $i \in \{1,\dots,a\}$.
We denote the set of $a \times b$ binary matrices by $\binmat{a \times b}$. 
For $X \in \binmat{a \times b}$, we define the set of matrices with the same sparsity pattern of $X$ as 
\begin{equation*}
    \sparse(X) := \{ Y \in \real{a\times b}\:|\: Y_{i,j}=0 \quad\forall i,j \:\st\: X_{i,j} = 0 \}\,.
\end{equation*}
With $\operatorname{card}(X)$, we represent the number of nonzero elements in the binary matrix $X$. 
For a matrix $Z\in\real{a\times b}$, $X := \struct(Z) \in \binmat{a\times b}$ if 
\begin{equation*}
    X_{i,j} =
    \begin{cases}
        0 & \textnormal{ if }Z_{i,j} = 0, \\
        1 & \textnormal{ otherwise.}
    \end{cases}\,.
\end{equation*}
For $X,Y\in\binmat{a \times b}$, we say $X \leq Y$ if $X_{i,j} \leq Y_{i,j}$, $\forall i,j$.
The operators $\norm{\cdot}_F$ and $\norm{\cdot}_{2 \to 2}$ represent the Frobenius and induced 2-norm of a matrix, respectively.
Given a symmetric matrix $H\in\real{a\times a}$, we denote the largest eigenvalue of $H$ as $\lambda_{\max}(H)$.
For a square matrix $X$, we use the notation $X \succ 0$ $(X \succeq 0)$ to denote positive (semi-) definiteness. 
The operator $\otimes$ represents the Kronecker product.
Finally, $\operatorname{blkdiag}(A,B,C, \dots)$ represents the block-diagonal matrix with matrices $A, B, C, \dots $ on the diagonal.

\preprintswitch{
\emph{Remark:} For the sake of conciseness, the proofs of the results in this work are deferred to the Appendix of the extended manuscript~\cite{martinelli2023closing}.
}{}

\section{Problem formulation}\label{sec:problem_formulation}
We consider discrete-time linear time-varying dynamical systems described by the state-space equations:
\begin{equation}
\label{eq:dynamical_system}
x_{t+1} = A_t x_t + B_t u_t + E_tw_t \,,
\end{equation}
where $x_t \in \mathbb{R}^n$, $u_t \in \mathbb{R}^m$, and $w_t \in \mathbb{R}^n$ represent the system state, the control input, and an exogenous disturbance, respectively.
Motivated by the observation that it is often difficult to characterize the class of disturbances, we make no assumptions regarding the distribution or the nature of $w_t$ over time.
For simplicity, we assume $E_t = I$. The more general case where $E_t \neq I$ can be addressed using the methods described in~\cite{didier2022system}.

We consider the scenario where the system described by \eqref{eq:dynamical_system} is controlled over a finite horizon $T \in \mathbb{N}^+$, starting from an initial condition $x_0 \in \mathbb{R}^n$. 
For compactness, we define
\begin{gather*}
    \tf{x} := 
    \begin{bmatrix}
        x_0 \\
        x_1 \\
        \vdots \\
        x_{T-1}
    \end{bmatrix}
    , \:\:
        \tf{u} := 
    \begin{bmatrix}
        u_0 \\
        u_1 \\
        \vdots \\
        u_{T-1}
    \end{bmatrix}
    , \:\:
        \ttf{\delta}
    := 
    \begin{bmatrix}
        x_0 \\
        w_0 \\
        \vdots \\
        w_{T-2}
    \end{bmatrix}
    = 
    \begin{bmatrix}
        x_0 \\
        \tf{w}
    \end{bmatrix}\,.
\end{gather*}
At each time instant $t$, we denote the input $u_t$ produced by a controller $\pi_t$ as $u_t = \pi_t (x_0, \ldots, x_t)$.
For a policy $\ttf{\pi} := [\pi_0^\top(x_0)\dots \pi_{T-1}^\top(x_0,\dots, x_{T-1})]^\top$ and a disturbance $\ttf{\delta}$, the incurred cost is defined as
\begin{equation}\label{eq:J_quadratic_cost}
J(\ttf{\delta}, \ttf{\pi})
:=
\begin{bmatrix}
\mathbf{x}^\top & \mathbf{u}^\top
\end{bmatrix}
\mathbf{C}
\begin{bmatrix}
\mathbf{x} \\
\mathbf{u}
\end{bmatrix}\,.
\end{equation}
where the matrix $\tf{C} \succeq 0$ assigns different weights to the states and input signals at different time instants.
Notice that it is intractable to cast an optimization program over the class of all general policies $\ttf{\pi}$. Motivated by their optimality for centralized linear quadratic control and their tractability properties, we focus on linear feedback policies of the form 
\(\mathbf{u} = \mathbf{K} \mathbf{x}\,,\)
where $\mathbf{K} \in \mathbb{R}^{mT \times nT}$ is a lower block-triangular matrix due to causality\footnote{
An affine policy $\tf{u} = \tf{K}\tf{x} + \tf{g}$ can be also considered augmenting the state as $\Tilde{x}_t := \begin{bmatrix} x_t^\top & 1\end{bmatrix}^\top$.
Thus, we will focus on linear feedback policies, without loss of generality.
}.
To highlight the dependency of $\mathbf{K}$ in~\eqref{eq:J_quadratic_cost}, we will use the notation $J(\ttf{\delta},\mathbf{K})$.

This paper focuses on large-scale systems, where each controller has only access to partial sensor information. 
We represent this condition using the constraint
\begin{equation*}
    \tf{K} \in \mathcal{S} \,,\quad \mathcal{S} = \sparse(\tf{S})\,,
\end{equation*}
where $\tf {S} \in\binmat{mT \times nT}$ describes the spatio-temporal information of the system, in the sense that $\tf{S}$ describes which scalar control input depends on which scalar state, and at which time instant. 
We will refer to $\tf{S}$ as ``sparsity matrix''. We always assume $\tf{S}$ to be lower block-diagonal to ensure the causality of the control policy.

\subsection{Spatial regret}

The cost function~\eqref{eq:J_quadratic_cost} depends on the realization of the unknown disturbances $\ttf{\delta}$. Hence, finding a sparse controller $\tf{K}$ minimizing~\eqref{eq:J_quadratic_cost} for any~$\ttf{\delta}$ is an ill-posed problem. 
To remove the explicit dependency of the cost on $\ttf{\delta}$, the $\H2$ paradigm optimizes the expected performance under the assumption of stochastic disturbances with zero average and finite second moment, while the $\Hinfty$ setup focuses on optimizing with respect to the worst-case disturbance realization. However, both setups are unlikely to hold in large-scale systems. The presence of unmodeled dynamics often leads to non-stochastic uncertainties, and disturbances are frequently localized to specific subsystems rather than representing a worst-case scenario.
In this work, we propose an alternative controller performance metric that is tailored to large-scale scenarios.

Suppose, to have an ideal sparsity matrix $\tf{\hat{S}}$ that is denser than the real control sparsity structure $\mathcal{S}$, i.e.,  $\mathcal{S} \subset \hat{\mathcal{S}}$, with ${\mathcal{S}}= \sparse(\tf{{S}})$, and $\hat{\mathcal{S}}= \sparse(\tf{\hat{S}})$.
Then, a controller $\tf{\hat{K}} \in \hat{\mathcal{S}}$ would have more sensor information to use, and it would be able to detect and counteract localized disturbances more promptly, regardless of the location where they may hit.
Our idea is to promote the design of sparse controllers that imitate, in hindsight, the behavior of denser ones.
To do so, first define the error between the cost of the controller $\tf{K}\in {\mathcal{S}}$ and $\tf{\hat{K}}\in \hat{\mathcal{S}}$ for a given $\ttf{\delta}$ as
\begin{equation}\label{eq:def_error_cost_e}
    e(\ttf{\delta}, \tf{K},\tf{\hat{K}}) := J(\ttf \delta, \tf K) - J (\ttf \delta, \tf{\hat{K}})\,.
\end{equation}
Then, we introduce a new metric that we call \textit{Spatial Regret} quantifying the worst-case scenario of~\eqref{eq:def_error_cost_e} as 
\begin{equation}\label{eq:def_spatial_regret_cost}
    \spreg ( \tf K , \tf{\hat{K}} ) :=
    \max_{\lVert \ttf \delta \rVert_{2} \leq 1}
   e(\ttf{\delta}, \tf{K},\tf{\hat{K}}) \,.
\end{equation}
Finally, the minimization problem we want to solve is
\begin{subequations}
\label{eq:minimize_spatial_regret_generic}
\begin{alignat}{2}
&\!\min_{\tf{K}}  &\quad& \spreg( \tf K , \tf{\hat{K}} )\\
&~\st                &     & 
\tf{K} \in \mathcal{S}\,, \quad \mathcal{S} \subset \hat{\mathcal{S}}\,. \label{eq:sparsity_constraints_regret}
\end{alignat}
\end{subequations}
From~\eqref{eq:def_spatial_regret_cost}, note that a non-positive spatial regret value implies that the oracle $\tf{\hat{K}}$ cannot improve the performance of the controller $\tf{K}$ to be designed for any disturbance $\ttf{\delta} \neq 0$. Therefore, \eqref{eq:def_spatial_regret_cost} is not \textit{well-posed }in the sense that it is meaningless to match the performance of a universally worse benchmark for any disturbance. An example demonstrating this point is reported \preprintswitch{in~{\cite{martinelli2023closing}[App. A]}}{in~\Cref{ap:numerical_example_bad_K_hat}}. 
Problem~\eqref{eq:minimize_spatial_regret_generic} presents two key challenges. First, we need to verify when it is well-posed, meaning that $\spreg(\cdot,\cdot)$ is lower bounded for any selected controller $\mathbf{K} \in \mathcal{S}$, given the oracle $\mathbf{\hat{K}}$.
Second, notice that~\eqref{eq:minimize_spatial_regret_generic} is composed of multiple nested optimization problems, making its solution intractable in this form. In \Cref{sec:convex_reformulation} we provide a convex approximation of~\eqref{eq:minimize_spatial_regret_generic}.
\begin{remark}
    Our $\spreg$ metric is inspired by \cite{sabag2021regret, goel2023regret, martin2022safe, didier2022system, martin2023guarantees, martin2023follow}, where centralized controllers are designed by minimizing regret with respect to a benchmark policy that has foreknowledge of future realization of $\ttf{\delta}$. Unlike these and related works rooted in online optimization, see e.g., \cite{ghai2022regret}, our approach introduces a spatial notion of regret instead of a temporal one.
    Studying the interplay between spatial and temporal notions of regret is left as an interesting direction for future research.
\end{remark}

\begin{remark}

In the case of a perfect oracle achieving zero cost for every disturbance realization, \eqref{eq:def_spatial_regret_cost} reduces to the objective of classic $\Hinfty$ controllers. However, the key strength of our metric emerges when considering imperfect oracles: the spatial regret metric allows mimicking the behavior of an unattainable policy that has additional sensor data (which we lack). This approach prioritizes mitigating disturbances where this additional information would be most beneficial, all while complying with the information sparsity of the system.
\end{remark}

\subsection{Review of convex design of distributed controllers}
We review results on the convex design of distributed controllers that are instrumental in solving the above challenges.
Let $\mathbf{Z}$ denote the block-downshift operator, namely a block-matrix with identity matrices along its first block sub-diagonal and zeros elsewhere.
Define the matrices $\mathbf{A} := \operatorname{blkdiag}(A_0, \dots , A_{T-2}, 0_{n \times n})$, and $\mathbf{B} := \operatorname{blkdiag}(B_0, \dots , B_{T-2}, 0_{n \times m})$.
Then, the state evolution of \eqref{eq:dynamical_system} can be represented compactly as
\begin{equation}\label{eq:state_evolution_compact_form}
    \tf x = \mathbf{Z} \mathbf{A} \, \tf x + \mathbf{Z} \mathbf{B} \, \tf u +\ttf{\delta} \,.
\end{equation}
Considering a control law $\tf u = \tf K \tf x $ and~\eqref{eq:state_evolution_compact_form}, it is straightforward to write the closed-loop maps $\ttf{\Phi}_x, \ttf{\Phi}_u$ from the disturbances $\ttf \delta$ to $\tf x$ and $\tf u$ as
\begin{equation} \label{eq:definition_Phi_x_Phi_u}
    \begin{bmatrix}
        \tf x \\
        \tf u
    \end{bmatrix}
    =
    \begin{bmatrix}
        \big(I - \mathbf{Z}(\mathbf{A} + \mathbf{B} \tf K)\big)^{-1}\\
        \tf K \big(I - \mathbf{Z}(\mathbf{A}+\mathbf{B} \tf K)\big)^{-1}
    \end{bmatrix}\ttf{\delta}
    =
    \begin{bmatrix}
        \ttf{\Phi}_x\\
        \ttf{\Phi}_u
    \end{bmatrix}
    \ttf{\delta} = \tf \Phi \ttf{\delta}.
\end{equation}
Here, $\ttf \Phi$ is the vertical stacking of $\ttf{\Phi}_x , \ttf{\Phi}_u$.
The closed-loop responses $\ttf{\Phi}_x, \ttf{\Phi}_u$ are lower block-diagonal due to causality. It is easy to verify that $\tf K$ and $\tf \Phi$ are linked through the relation $\tf{K} = h(\ttf{\Phi}) =  \ttf{\Phi}_u \ttf{\Phi}_x^{-1}$, with $h:\real{(m+n)T \times nT} \to \real{mT \times nT}$.
Also, as shown in~\cite{anderson2019system}, it can be proved that there exists a controller $\tf K$ such that \eqref{eq:definition_Phi_x_Phi_u} holds if and only if
\begin{equation}
\label{eq:achievability_constraint}
    (I - \mathbf{Z} \mathbf{A}) \ttf{\Phi}_x - \mathbf{Z} \mathbf{B} \ttf{\Phi}_u = I \,.
\end{equation}
With the introduction of the maps $\ttf{\Phi}_x, \ttf{\Phi}_u$, the cost $J(\ttf{\delta},\tf K)$ can be rewritten as  
\begin{equation}\label{eq:def_J_in_Phi}
    J (\ttf{\delta}, \tf K) 
    =
    \ttf{\delta}^\top
    \ttf{\Phi}^\top
    \mathbf{C}
    \,
    \ttf{\Phi}
    \,
    \ttf{\delta}\,.
\end{equation}
Moreover, as shown in~\cite{anderson2019system}, the classic $\H2$ and $\Hinfty$ control problems, can be reformulated as
\begin{align} 
    &\H2: \quad \mathbb{E}_{\ttf{\delta} \sim \mathcal{D}} [{J(\ttf {\delta},\tf{{K}})}] = 
    \norm{\tf{C}^{\frac{1}{2}} \tf{\Phi} \Sigma_{\ttf{\delta}} }_F^2\,,
    \label{eq:H2_cost_in_Phi}
    \\
    &\Hinfty: \quad  \max_{\lVert \ttf \delta \rVert_{2} \leq 1}(J(\ttf {\delta},\tf{{K}})) = 
    \norm{\tf{C}^{\frac{1}{2}} \tf{\Phi}}_{2 \to 2}^2\,,
    \label{eq:Hinfty_cost_in_Phi}
\end{align}
where $\mathcal{D}$ in \eqref{eq:H2_cost_in_Phi} denotes the probability distribution of $\bm{\delta}$, with zero mean and covariance $\Sigma_{\ttf{\delta}} \succeq 0$.
Thus, the costs~\eqref{eq:H2_cost_in_Phi} and \eqref{eq:Hinfty_cost_in_Phi} are convex in $\tf{\Phi}$. Nonetheless, the sparsity condition~\eqref{eq:sparsity_constraints_regret} is nonconvex in $\tf{\Phi}$. However, when the QI condition defined next holds, the sparsity constraint~\eqref{eq:sparsity_constraints_regret} can be rewritten linearly in~$\tf{\Phi}$~\cite{rotkowitz2005characterization}. 
 \begin{definition}\label{def:QI}
 Define $\tf G := (I - \mathbf{Z}\mathbf{A})^{-1}\mathbf{Z}\mathbf{B}$. 
A subspace $\mathcal{S} \subseteq \real{mT \times nT}$ is QI with respect to $\tf G$ if and only if \[
    \tf{K G K} \in {\mathcal{S}} \,, \quad \forall \tf K \in {\mathcal{S}}\,.
\]
 \end{definition}
If $\mathcal{S}$ is QI with respect to $\tf G$, then it is was shown~\cite{furieri2019unified} that
\begin{equation}\label{eq:QI_equivalence_K_sparse}
   \tf{{K}} \in {\mathcal{S}} \quad \Leftrightarrow \quad \tf{{\Phi}}_u \Gamma \in {\mathcal{S}}\,,
\end{equation}
where $\Gamma := I - \tf{ZA}$.
Thus, the QI condition enables us to search over all possible sparse controllers $\tf{K} \in \mathcal{S}$, allowing globally optimal minimization of any convex cost, such as~\eqref{eq:H2_cost_in_Phi}, \eqref{eq:Hinfty_cost_in_Phi}.
However, the QI condition may not be satisfied for every desired dynamical system or sparsity matrix. In~\cite{furieri2020sparsity}, a technique was introduced for deriving sparse controllers, even in cases where QI does not hold.
Specifically, the work \cite{furieri2020sparsity} establishes a method to compute, given $\mathcal{S}$, a matrix $\tf{V}_x \in \binmat{nT \times nT}$ such that
\begin{subequations}\label{eq:SI}
\begin{gather}
\ttf{\Phi}_u \Gamma \in \mathcal{S} \quad \textnormal{and} \quad \ttf{\Phi}_x  \Gamma\in \sparse(\tf{V}_x)\,,\label{eq:SI_definition}
\\
\Downarrow \nonumber
\\
\tf{K} = h(\tf\Phi) = \ttf{\Phi}_u \ttf{\Phi}_x^{-1} \in \mathcal{S}\,,\label{eq:SI_definition_result}
\end{gather}
\end{subequations}
where $\tf{\Phi}_x$, $\tf{\Phi}_u$ satisfy~\eqref{eq:achievability_constraint}.
The authors refer to~\eqref{eq:SI} as the Sparsity Invariance~(SI) condition.  A minimally restrictive choice of the binary matrix $\tf{V}_x$ complying with~\eqref{eq:SI} is given by~\cite[Alg. 1]{furieri2020sparsity}, which we report here as~\Cref{alg:Generation_Vx} for the sake of completeness.
\begin{algorithm}
\caption{Generation of $\tf{V}_x$~\cite[Alg. 1]{furieri2020sparsity}}\label{alg:Generation_Vx}
\begin{algorithmic}[1]
\Require $\tf{S}$ (System's Sparsity Matrix)
\State Initialize $\tf{V}_x = \tf{1}_{nT \times nT}$
\For{each $i = 1,\dots,mT$, $k=1,\dots,nT$}
    \If{$\tf{S}_{i,k} == 0$}
        \For{each $j = 1,\dots,nT$}
            \State \textbf{if} $\tf{S}_{i,j} == 1$ \textbf{then} $(\tf{V}_x)_{j,k} \gets 0$%
        \EndFor
    \EndIf
\EndFor
\end{algorithmic}
\end{algorithm}
In this paper, we denote the set of sparse controllers parametrized by SI as
\begin{equation}
\label{eq:Theta}
          \tf{\Theta} := \{h(\tf{\Phi}) : \: \eqref{eq:achievability_constraint},\:\: \eqref{eq:SI_definition} \}\,.
\end{equation}
If $\tf{V}_x$ is designed according to \Cref{alg:Generation_Vx}, then we have $\tf{\Theta}\subseteq \mathcal{S}$.

\section{Main results}\label{sec:main_results}
In this section, we address the two key challenges of problem~\eqref{eq:minimize_spatial_regret_generic} regarding the well-posedness of the metric and its convex reformulation.
First, we establish that if the oracle $\tf{\hat{K}} \in \hat{\mathcal{S}}$ is optimal according to $\H2$ or $\Hinfty$ criteria, then $\spreg$ is well-posed in the sense that imitating its behavior is advantageous for some $\ttf{\delta}$.
Second, we demonstrate that the QI condition plays a crucial role in the design of the oracle $\tf{\hat{K}}$ for enabling a convex synthesis of the spatial regret optimal policy $\tf{{K}}$.

\begin{proposition}\label{prop:well_posed_regret}
Assume an oracle $\tf{\hat{K}}$ is derived by solving the optimization problem
\begin{equation}\label{eq:def_oracle_problem}
\min_{\tf{\hat{K}}} f(\tf{\hat{K}}) \quad \st \quad \tf{\Hat{K}} \in \hat{\mathcal{S}}\,, \quad \mathcal{S} \subset \hat{\mathcal{S}}
\end{equation}
where $f(\tf{\hat{K}})$ is
$\mathbb{E}_{\ttf{\delta} \sim \mathcal{D}} [{J(\ttf {\delta},\tf{\hat{K}})}]$ 
or
$\max_{\lVert \ttf \delta \rVert_{2} \leq 1}(J(\ttf {\delta},\tf{\hat{K}}))$.
Then, $\spreg (\tf{K}, \tf{\hat{K}}) \geq 0$ for any $\tf K \in \mathcal{S}$.
\end{proposition}

\preprintswitch{}{The proof is reported in~\Cref{ap:proof_Proposition_choice_Oracle}.}
\Cref{prop:well_posed_regret} sheds light on two possible design criteria. One could follow to synthesize a denser controller $\tf{\hat{K}} \in \hat{\mathcal{S}}$ that is guaranteed to be informative, in the sense that no other controller $\tf{K} \in \mathcal{S}$ can outperform it for all possible realizations of $\ttf{\delta}$.
Despite these advancements, how to obtain an oracle $\tf{\hat{K}}$ with sparsity constraints $\tf{\hat{S}}$ by solving~\eqref{eq:def_oracle_problem} remains still a non-trivial challenge. The next section addresses this crucial aspect in detail.

\begin{remark}
With an unconstrained noncausal benchmark, obtaining an optimal benchmark for any $\ttf{\delta}$ is possible, as shown in \cite{martin2022safe, hassibi1999indefinite}. Therefore, formulations that minimize worst-case regret with a noncausal oracle are always well-posed. Our result in \Cref{prop:well_posed_regret} reaffirms that spatial regret preserves this property even when introducing sparsity constraints in the oracle design.
\end{remark}

\subsection{Synthesis of the  oracle \texorpdfstring{$\tf{\hat{K}}$}{KHat}}
\label{subsec:oracle_synthesis}
Thanks to \Cref{prop:well_posed_regret}, we have demonstrated there always exist choices for the oracle $\tf{\hat{K}}$ that ensure the well-posedness of problem \eqref{eq:minimize_spatial_regret_generic}. For example, an optimal centralized $\H2$ or $\Hinfty$ oracle $\tf{\hat{K}}$, is a valid choice. 
However, a key challenge remains how to effectively synthesize the distributed controllers  
$\tf{K}$ and $\tf{\hat{K}}$ while guaranteeing well-posedness of the spatial regret.
Since~\eqref{eq:SI} yields linear constraints on $\tf{\Phi}$ that come with tightness guarantees and comply with any desired information structure, a seemingly natural approach would be to exploit \Cref{alg:Generation_Vx} to synthesize both $\tf{V}_x$ and $\tf{\hat{V}}_x$, satisfying the SI condition \eqref{eq:SI} for $\mathcal{S}$ and $\hat{\mathcal{S}}$, respectively.
It must be pointed out that these steps might fail to obtain a well-posed spatial regret metric, in general. An example illustrating this point is detailed \preprintswitch{in~\cite[App. C]{martinelli2023closing}}{in~\Cref{ap:example_QI_is_sufficient}}.
Nonetheless, in the following theorem, we reveal the crucial role of the QI condition in designing an informative oracle $\tf{\hat{K}}$ in conjunction with SI for the synthesis of the spatial regret optimal policy for any $\mathcal{S}$.

\begin{theorem}\label{th:well_posed_with_QI}
Let $\hat{\mathcal{S}}$ be QI with respect to $\tf{G}$, with ${\mathcal{S}}\subset \hat{\mathcal{S}}$. Additionally, denote with $\tf{V}_x$ and $\tf{\hat{V}}_x$ the binary matrices computed using~\Cref{alg:Generation_Vx} based on ${\mathcal{S}}$ and $\hat{\mathcal{S}}$, respectively.
Assume an oracle $\tf{\hat{K}} = h(\tf{\hat{\Phi}})$ is derived through the optimization problem
\begin{equation}\label{eq:oracle_min_problem_in_Phi}
\min_{\tf{\hat{\Phi}}} f(\tf{\hat{\Phi}})
\:\:
\st 
\:
\eqref{eq:achievability_constraint},\:\:
\tf{\Hat{\Phi}}_u \Gamma \in \hat{\mathcal{S}}, 
\:
\tf{\Hat{\Phi}}_x \Gamma \in \sparse(\tf{\hat{V}}_x),
\end{equation}
where the function $f(\tf{\hat{\Phi}})$ can be either~\eqref{eq:H2_cost_in_Phi} or \eqref{eq:Hinfty_cost_in_Phi} and $\Gamma :=(I - \tf{ZA})$.
Then, $\spreg (\tf{K}, \tf{\hat{K}}) \geq 0$ for any $\mathbf{K} \in \tf{\Theta}$.
\end{theorem}

\preprintswitch{}{The proof is reported in~\Cref{app:proof_th_well_posed_with_QI}.}
With~\Cref{th:well_posed_with_QI}, we proved that if the oracle structure satisfies the QI condition, then the well-posedness of the spatial regret metric is preserved using the SI condition~\eqref{eq:SI} to enforce both sparsity constraints over the oracle $\tf{\hat{K}}$ and then the actual controller $\tf{K}$.

\begin{remark}
{Given the sparsity matrix $\tf{S}$, one can obtain its closest QI superset $\tf{\hat{S}}_{QI}$ such that 
\begin{equation}\label{eq:nearest_QI_problem}
\begin{aligned}
{\tf{\hat{S}}_{QI}} :=&\operatorname*{argmin}_{\tf{S}^*} &&\operatorname{card}(\tf{S}^* - \tf{S})\\
&~~~\st  &&
\mathcal{S} \subseteq {\mathcal{S}^*}, \quad  {\mathcal{S}^*}= \sparse(\tf{S}^*) ,
\\
&&&
\mathcal{S}^* \textnormal{ is QI w.r.t. } \tf{G}.
\end{aligned}
\end{equation}
In~\cite{rotkowitz2011nearest} the authors proposed a method to obtain efficiently $\tf{\hat{S}}_{QI}$ in a finite number of steps. While results hold for any QI subspace $\hat{\mathcal{S}} \supset \mathcal{S}$, the method \eqref{eq:nearest_QI_problem} can always be utilized to generate  $\hat{\mathcal{S}}$.}
\end{remark}
\begin{remark}\label{rk:difference_cardinality_reg_oracle}
It is important to note that the ``distance'' in terms of cardinality between the two sparsity matrices $\tf{S}$ and $\tf{\hat{S}}$ plays an important role in obtaining well-performing controllers. 
When $\operatorname{card}(\tf{S})$ is approximately equal to $\operatorname{card}(\tf{\hat{S}}$), the controller $\tf{K}$ may struggle to learn adequately, leading it to imitate the oracle $\tf{\hat{K}}$. Conversely, when the two structures are significantly different, with $\operatorname{card}(\tf{S}) \ll \operatorname{card}(\tf{\hat{S}})$, the behavior of the oracle may be too challenging for the actual controller to follow, resulting in poor performance. Choosing the nearest QI superset has been demonstrated to yield heuristically good performance, particularly when the actual matrix $\tf{G}$ is highly sparse.
\end{remark}

\subsection{Convex reformulation of the spatial regret problem}\label{sec:convex_reformulation}
Here, we tackle the challenge of obtaining a convex approximation of the $\spreg$ minimization problem~\eqref{eq:minimize_spatial_regret_generic}.

\begin{proposition}\label{prop:dual_regret}
    Let $\tf{\hat{K}} = h(\tf{\hat{\Phi}}) \in \sparse(\tf{\hat{S}})$ be a solution of \eqref{eq:oracle_min_problem_in_Phi}, where $\tf{\hat{S}}$ is QI with respect to $\tf{G}$. Consider the following convex optimization problem
    \begin{subequations}\label{eq:regret_problem_dual}
        \begin{align}
        &  \underset{\tf{\Phi}, \lambda}{\min} &&\lambda \label{eq:regret_problem_dual_cost}\\
        &\st && \eqref{eq:achievability_constraint}\,,\: \eqref{eq:SI_definition}\,,\nonumber\\
        & {}&&
        \begin{bmatrix}
            I & \mathbf{C}^{\frac{1}{2}} \tf{\Phi} \\
            \tf{\Phi}^\top \mathbf{C}^{\frac{1}{2}} & \lambda I + \tf{\hat{\Phi}}^\top \mathbf{C} \tf{\hat{\Phi}}
        \end{bmatrix}
        \succeq 0\,,\label{eq:regret_problem_dual_pos_def_constraint}
        \end{align}
    \end{subequations}
    where $\tf{V}_x$ in \eqref{eq:SI_definition} is designed according to \Cref{alg:Generation_Vx}. Let the set of parametrized controllers $\tf{\Theta} \subseteq \mathcal{S}$ be defined as per \eqref{eq:Theta}. Finally, let $\bm{\Phi}^\star$ denote an optimal solution to \eqref{eq:regret_problem_dual} and $\mathbf{K}^\star = h(\bm{\Phi}^\star)$ the corresponding controller. Then:
    \begin{enumerate}
        \item $\spreg(\mathbf{K},\tf{\hat{K}})\geq 0$  for any $\mathbf{K} \in \bm{\Theta}$,
        \item  $\spreg(\mathbf{K}^\star,\tf{\hat{K}}) \leq\spreg(\mathbf{K},\tf{\hat{K}})$ for all $\tf{K}\in \tf{\Theta}$,
         \item If $\mathcal{S}$ is QI with respect to $\mathbf{G}$, then  $\mathbf{K}^\star$ is a globally optimal solution to problem \eqref{eq:minimize_spatial_regret_generic}. 
    \end{enumerate}
\end{proposition}
\preprintswitch{}{The proof is reported in~\Cref{app:proof_prop_dual_regret}.}
For the sake of clarity, we summarize the steps of our controller design method in~\Cref{alg:calculus_K}.
\vspace*{-5pt}
\begin{algorithm}
\caption{Controller synthesis for $\spreg$}\label{alg:calculus_K}
\begin{algorithmic}[1]
\Require $\tf{S}$ (System's Sparsity Matrix)
\State {Choose a QI subspace $\hat{\mathcal{S}}\supseteq\mathcal{S}$, for instance, using~\eqref{eq:nearest_QI_problem}}.
\State {Compute $\tf{V}_x$ and $\tf{\hat{V}}_x$ with \Cref{alg:Generation_Vx}, starting from $\mathbf{S}$ and $\hat{\mathbf{S}}$, respectively}.
\State Derive the oracle controller $\tf{\hat{K}}$ by solving~\eqref{eq:oracle_min_problem_in_Phi}.
\State Obtain the $\spreg$ controller $\tf{{K}}$ through~\eqref{eq:regret_problem_dual}.
\end{algorithmic}
\end{algorithm}
\vspace{-0.35cm}
\section{Numerical results}\label{sec:numerical_results}

We conduct a comparative analysis between distributed controllers minimizing spatial regret metrics and traditional $\tf{K}_{\H2}$ and $\tf{K}_{\Hinfty}$ ones, all utilizing the same sparsity structure $\mathcal{S}$ and synthesized using the SI condition~\eqref{eq:SI}. We use a multi-mass spring-damper system consisting of $10$ masses within a time window of $T = 30$ and a sampling time of $T_s = \SI{0.5}{\second}$. For full model details and simulation settings, please refer \preprintswitch{to~\cite[App. F]{martinelli2023closing}}{to~\Cref{ap:implementation_details}}.\footnote{The code used in this work is accessible at~\url{https://github.com/DecodEPFL/SpRegret}.}

\begin{figure}[t]
    \centering
    \includegraphics[width=0.4\textwidth]{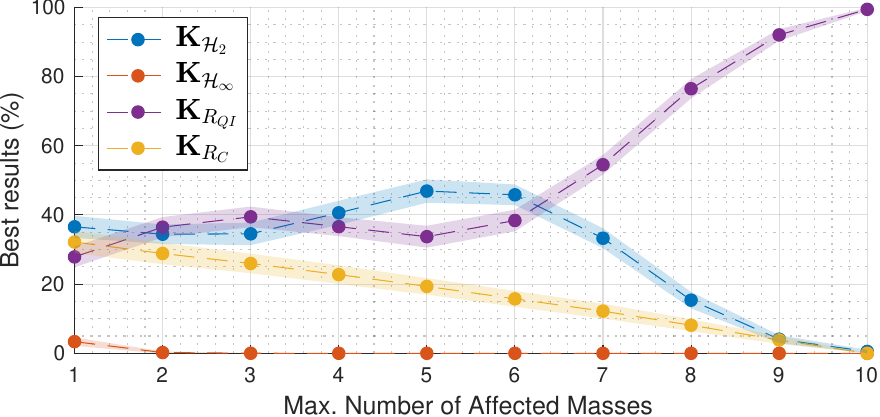}
    \caption{Percentage of times a control policy outperforms the remaining three as a function of the maximum number of masses affected by uniformly distributed disturbances in the interval $[-0.5,1]$ in a system of $10$ masses. Shaded areas around the dotted lines represent the $95\%$ confidence interval around the corresponding mean values.}
    \label{fig:random_affected_cars}
\end{figure}

\begin{figure}[t]
    \centering
        \includegraphics[width=0.4\textwidth]{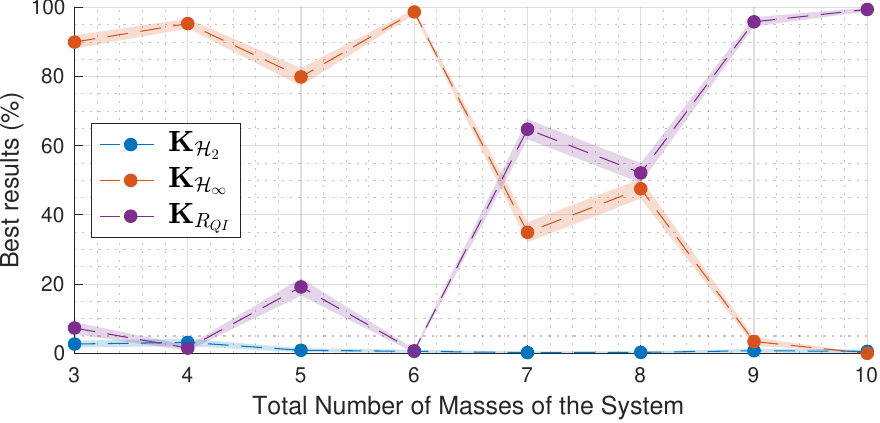}
    \caption{
    Percentage of times a control policy outperforms the remaining two as a function of the number of masses constituting the large-scale systems when all the masses are affected by uniformly distributed disturbances in the interval $[-0.5,1]$. Shaded areas around the dotted lines represent the $95\%$ confidence interval around the corresponding mean values.
    }
    \label{fig:increasing_number_masses}
\end{figure}

We design two $\spreg$ controllers: $\tf{K}_{R_{QI}}$ and $\tf{K}_{R_{C}}$. The former employs an oracle with sparsity determined by the nearest QI superset of $\tf{S}$. The latter imitates a centralized oracle, which is guaranteed to be QI by definition.
We perform two distinct experiments, the results of which are depicted in Fig.~\ref{fig:random_affected_cars} and Fig.~\ref{fig:increasing_number_masses}. In both cases, our main focus is to determine the percentage of times in which each controller yields a better (i.e., smaller) cost $J(\cdot)$ compared to the other three.

In the first experiment, we aim to replicate the unpredictability and uncertain nature of disturbances by simulating perturbations $\ttf{\delta}$, drawn from a non-centered uniform distribution $\mathcal{U}[-0.5, 1]$. These disturbances are then applied to a random number of masses within the system. The number of affected masses is also drawn uniformly, within the interval $[1,N]$. We gradually increase the value of $N$ up to the total number of interconnected subsystems (in this case $10$). 
The results are presented in~Fig.~\ref{fig:random_affected_cars}. 
The percentage value of times each policy yields better performance is computed over $1000$ realizations of $\ttf{\delta}$ and, to obtain the $95\%$ confidence interval of these measurements, we iterated this step $100$ times, for a total of $10^5$ experiments.
As expected, when the number of affected masses is small, the resulting disturbance is zero for most of the agents, aligning with the $\H2$ hypothesis on a distribution $\mathcal{D}$ centered around $0$. However, as the number of affected masses increases, the cumulative effect of non-standard $\ttf{\delta}$ on the overall system becomes more pronounced, thus favoring the performance of the controller $\tf{K}_{R_{QI}}$. 
In the table below, we report the average cost of the controller $\tf{K}_{R_{QI}}$ over $10^5$ experiments and the relative increase in cost for the other controllers compared to $\tf{K}_{R_{QI}}$ for an increasing number of masses.
\vspace*{-0.4cm}
\begin{table}[H]
    \centering
    \rowcolors{1}{}{lightgray}
    \begin{tabular}{c|c|ccc}
        N.Masses & Av. cost $\tf{K}_{R_{QI}}$ & $\tf{K}_{\H2}$ & $\tf{K}_{\Hinfty}$ & $\tf{K}_{R_{C}}$\\
        \hline
        $1$ &  $14.20$  & $+7.39\%$  & $+1.35\%$ & $+0.44\%$\\
        $5$ &  $30.21$  & $+13.43\%$ & $+3.32\%$ & $+1.25\%$\\
        $10$ & $39.35$  & $+43.82\%$ & $+4.32\%$ & $+1.20\%$\\
        \hline
    \end{tabular}
\end{table}
\vspace*{-0.4cm}
As we can observe, $\tf{K}_{R_{QI}}$ consistently achieves lower average control costs. Notably, the cost advantage of $\tf{K}_{QI}$ over other controllers increases with system complexity (number of masses). For instance, with $10$ masses, $\tf{K}_{R_{QI}}$ outperforms the $\tf{K}_{\H2}$ controller by over $43\%$, while just improving by $4\%$ and $1\%$ compared to $\tf{K}_{\Hinfty}$ and $\tf{K}_{R_{C}}$, respectively. This demonstrates the effectiveness of our spatial regret-based approach in achieving efficient control, particularly for larger and potentially more challenging systems.

In the second experiment, we evaluate the performance improvement of controllers minimizing spatial regret as the sparsity structure of the large-scale system becomes more and more distributed.
For this reason, we choose a QI benchmark as it strikes a favorable balance between imitating a sparse structure and a centralized one.
We consider the same multi-mass spring damper system of the first experiment with a progressively increasing number of masses, from $3$ to $10$. For each new configuration, we synthesize again all the controllers ($\tf{K}_{\H2}$, $\tf{K}_{\Hinfty}$, $\tf{K}_{R_{QI}}$). Then, we simulate the effects of perturbations $\ttf{\delta}$ drawn from a non-centered uniform distribution $\mathcal{U}[-0.5,1]$ applied to all the masses with the same amount of experiments as in the previous study case. The results are shown in~Fig.~\ref{fig:increasing_number_masses}. 
It is evident that for a small number of systems affected by noise, $\tf{K}_{\Hinfty}$ outperforms the other control policies. However, as the size of the system increases, and consequently its sparsity, $\tf{K}_{R_{QI}}$ exhibits superior performance, highlighting its capacity to leverage information from the ideal oracle.
\section{Conclusions}
In this work, we aimed to design and synthesize distributed controllers for large-scale linear dynamical systems, affected by localized and highly heterogeneous disturbances. To do so, we first introduced the novel metric $\spreg$. Then, we demonstrated its well-posedness when the oracle satisfies the QI condition. 
Finally, we provided a convex formulation for designing regret-optimal controllers with arbitrary sparsity structures, optimizing them to close the performance gap to an ideal QI subspace that encodes richer information for the control policy.
To illustrate the real-world relevance of our approach, we provided numerical examples involving a multi-mass spring-damper system that requires to be controlled in the presence of non-standard disturbances.
Through comparisons with classic $\H2$ and $\Hinfty$ policies, our results showcased the superior performance of $\spreg$ controllers in handling disturbances that may target large-scale distributed systems.

Future research will explore methods for designing constrained benchmarks that ensure non-negative regret and improved performance over $\H2$ and $\Hinfty$ for user-defined disturbance classes. Possible new directions will include investigating how to automatically select the sparsity structure of the oracle tailored to the problem to guarantee better performance, extend our results to the infinite-horizon case, and consider larger and more complex applications to showcase the potentiality of our novel metric.

\bibliography{bibliography.bib}

\preprintswitch{}{
\appendix

\subsection{Example of oracle not guaranteeing well-posedness}\label{ap:numerical_example_bad_K_hat}
Consider the following system
\begin{equation*}
    x_{t+1} = 
\begin{bmatrix}
1 & 0 \\
-1& 1
\end{bmatrix}
 x_t + I \:u_t +  I \:w_t,
\end{equation*}
and two different static controllers 
\begin{equation*}
K =- 
\begin{bmatrix}
1 & 0 \\
0 & 1
\end{bmatrix}
 ,\quad 
\hat{K} =+
\begin{bmatrix}
1 & 0 \\
1 & 1 
\end{bmatrix}
\end{equation*}
applied at each time instant. This corresponds to $\tf{K} = I_{T} \otimes K$ and $\tf{\hat{K}} = I_{T} \otimes \hat{K}$, where $I_{T} := I \in \real{T \times T}$. Assume $\tf{C} = I $. Then it can be verified that  $J (\ttf{\delta},\tf{K})$ converges to a finite number for any $\ttf{\delta}$, while $J(\ttf{\delta},\tf{\hat{K}})$ diverges to $+\infty$ with $T \to \infty$. As a result, $\spreg(\tf{K},\tf{\hat{K}}) = \max_{\norm{\ttf{\delta}}_2\leq 1 } J (\ttf{\delta},\tf{K}) - J(\ttf{\delta},\tf{\hat{K}})$  will be negative and diverge to  $-\infty$. This is because the matrix $(A+{\hat{K}})$ is not Schur (stable), implying that the closed loop system is unstable for any $\ttf{\delta} \neq 0$.

\subsection{Proof of Proposition~\ref{prop:well_posed_regret}}\label{ap:proof_Proposition_choice_Oracle}
Given $\tf{\hat{K}}$ from \eqref{eq:def_oracle_problem}, it always holds that
    \begin{equation}\label{eq:th_1_step_1}
        \forall \tf{\tilde{K}} \in \hat{\mathcal{S}} \rightarrow \exists \ttf {\delta^*}: J(\ttf {\delta^*},\tf{\Tilde{K}}) \geq 
        J(\ttf{\delta^*},\tf{\hat{K}})\,.
    \end{equation}
     This can be proved by contradiction. Indeed, if \eqref{eq:th_1_step_1} is false, then 
    \begin{equation*}
        \exists \tf{\Tilde{K}} \in \hat{\mathcal{S}} \rightarrow \forall \ttf{\delta}: \, J(\ttf {\delta},\tf{\Tilde{K}}) < 
        J(\ttf{\delta},\tf{\hat{K}})\,.
    \end{equation*}
    Since the above should hold for every $\ttf{\delta}$, then it must also hold for its average realization and worst-case ones, that is,
    $\mathbb{E}_{\ttf{\delta}\sim \mathcal{D}} [{J(\ttf {\delta},\tf{\Tilde{K}})}]< \mathbb{E}_{\ttf{\delta}\sim \mathcal{D}} [J(\ttf {\delta},\tf{\hat{K}})]
    $, and $\max_{\lVert \ttf \delta \rVert_{2} \leq 1}(J(\ttf{\delta},\tf{\Tilde{K}}))<\max_{\lVert \ttf \delta \rVert_{2} \leq 1}(J(\ttf{\delta},\tf{\hat{K}}))$. This goes against the definition of $\tf{\hat{K}}$ from \eqref{eq:def_oracle_problem}, whether obtained minimizing $\mathbb{E}_{\ttf{\delta}\sim \mathcal{D}}[J(\ttf \delta, \tf{\hat{K}})]$ or $\max_{\lVert \ttf \delta \rVert_{2} \leq 1}(J(\ttf \delta, \tf{\hat{K}}))$. For this reason, \eqref{eq:th_1_step_1} must hold.
    By \eqref{eq:th_1_step_1}, it immediately follows that
    \begin{equation}\label{eq:last_step_firts_preposition}
        \forall \tf{K} \in \mathcal{S} \subset\hat{\mathcal{S}} \rightarrow \exists \ttf {\delta}: J(\ttf {\delta},\tf{K}) - 
        J(\ttf{\delta},\tf{\hat{K}})\geq0\,.
    \end{equation}
    Hence, using definition~\eqref{eq:def_spatial_regret_cost} of spatial regret, \eqref{eq:last_step_firts_preposition} shows that $\spreg ( \tf K , \tf{\hat{K}} )\geq 0 $ for any $\tf K \in \mathcal{S}$.

\subsection{Example that using SI to design both \texorpdfstring{$\tf{{K}}$}{K}, \texorpdfstring{$\tf{\hat{K}}$}{KHat} does not guarantee well-posedness}\label{ap:example_QI_is_sufficient}
Let $\tf{\Delta} := \struct(\tf{G}) = I$ and 
\begin{equation*}
        \struct(\mathcal{S}) = 
    \begin{bmatrix}
        1 & 0 & 0\\
        1 & 1 & 0\\
        0 & 0 & 1
    \end{bmatrix},\:\:
        \struct(\hat{\mathcal{S}}) = 
    \begin{bmatrix}
        1 & 0 & 0\\
        1 & 1 & 0\\
        0 & 1 & 1
    \end{bmatrix}.
\end{equation*}
It is easy to verify that $\hat{\mathcal{S}}$ is not QI with respect to $\tf{G}$.
Then, the matrices $\tf{V}_x$ and $\tf{\hat{V}}_x$ obtained using~\Cref{alg:Generation_Vx} are
\begin{equation*}
    \tf{V}_x =
    \begin{bmatrix}
     1  &   0  &   0\\
     1  &   1   &  0\\
     0 &    0  &   1
    \end{bmatrix}
    \,,\:\:
    \tf{\hat{V}}_x =
    \begin{bmatrix}
     1   &  0  &   0\\
     0  &   1   &  0\\
     0  &   1   &  1
    \end{bmatrix}.
\end{equation*}
It is clear that $\tf{V}_x \nleq \tf{\hat{V}}_x$, and thus, $\tf{\Theta} \nsubseteq \tf{\hat{\Theta}}$. Consequently, the $\spreg$ is not guaranteed to be well-posed in this case.

\subsection{Proof of Theorem~\ref{th:well_posed_with_QI}}\label{app:proof_th_well_posed_with_QI}
The proof follows is structured in two steps. First, we establish that if $\hat{\mathcal{S}}$ satisfies the QI condition, then it implies $\tf{\Theta} \subseteq \tf{\hat{\Theta}}$, where $\tf{\Theta}$ is defined in~\eqref{eq:Theta} and
\begin{equation*}
    \tf{\hat{\Theta}} := \{h(\tf{\hat{\Phi}}) : \eqref{eq:achievability_constraint},\:\: \tf{\hat{\Phi}}_u \Gamma \in \hat{\mathcal{S}},\:\:\tf{\hat{\Phi}}_x \Gamma \in \sparse(\tf{\hat{V}}_x)\}\,.
\end{equation*}
Second, we demonstrate that the requirements of~\Cref{prop:well_posed_regret} are satisfied and $\spreg (\tf{K}, \tf{\hat{K}})\geq0$ for every $\tf{K} \in \mathcal{S}$. 

We proceed to show that $\tf{\Theta} \subseteq \tf{\hat{\Theta}}$.
Since $ \mathcal{S} \subset \hat{\mathcal{S}}$ by construction, then $\tf{\Phi}_u \Gamma \in \mathcal{S} \Rightarrow \tf{\Phi}_u \Gamma \in \hat{\mathcal{S}}$.
From~\eqref{eq:achievability_constraint}, we can rewrite $\tf{\Phi}_x$ as
\begin{equation}\label{eq:proof_wp_QI_first_step}
    \tf{\Phi}_x = (I - \tf{ZA})^{-1} + (I - \tf{ZA})^{-1}\tf{ZB}\tf{\Phi}_u\,.
\end{equation}
Right-multiplying both sides of~\eqref{eq:proof_wp_QI_first_step} by $\Gamma$ and using the definition of $\tf{G} := (I - \tf{ZA})^{-1}\tf{ZB}$, we obtain that 
\(
    \tf{\Phi}_x \Gamma= I + \tf{G}\tf{\Phi}_u \Gamma
\).
Let $\tf{\Delta} := \struct(\tf{G})$. Knowing that $\tf{G} \in \sparse(\tf{\Delta})$ and $\tf{\Phi}_u \Gamma \in \sparse(\mathcal{S})$, then it must hold that $I + \tf{G}\tf{\Phi}_u \Gamma \in \sparse(I + \tf{\Delta}\mathcal{S})$. This means that
\begin{equation*}
    \tf{\Phi}_x \Gamma 
    \in \sparse(I + \tf{\Delta}\mathcal{S}) \subseteq \sparse(I + \tf{\Delta}\hat{\mathcal{S}})\,.
\end{equation*}
From~\cite[Th. 4]{furieri2020sparsity}, we know that if $\hat{\mathcal{S}}$ satisfies the QI condition with respect to $\tf{G}$, then $\sparse(I + \tf{\Delta}\hat{\mathcal{S}}) \subseteq \sparse(\tf{\hat{V}}_x)$, which in turn implies that $\tf{\Phi}_x \Gamma \in \sparse(\tf{\hat{V}}_x)$.
Having established that $\tf{\Theta} \subseteq \tf{\hat{\Theta}}$, it follows that if a controller $\tf{K} \in \tf{\Theta}$, then also  $\tf{K} \in \tf{\hat{\Theta}}$.

Regarding the second step of the proof, $\tf{\hat{K}} = h(\tf{\hat{\Phi}})$ is derived through~\eqref{eq:oracle_min_problem_in_Phi}, which is equivalent to solving~\eqref{eq:def_oracle_problem} thanks to the QI condition. Finally, $\tf{\Phi}$ is such that~\eqref{eq:SI_definition} holds, thus $\tf{K} = h(\tf{\Phi}) \in \mathcal{S}$ thanks to~\eqref{eq:SI} and $\spreg (\tf{K}, \tf{\hat{K}})\geq0$ for~\Cref{prop:well_posed_regret}.

\subsection{Proof of Proposition~\ref{prop:dual_regret}}\label{app:proof_prop_dual_regret}

1) The first statement directly follows by Theorem~\ref{th:well_posed_with_QI} because $\bm{\Phi}$ complies with the constraints defining the set $\tf{\Theta}$. 2) For every $\mathbf{K} \in \tf{\Theta}$, it is true that
\begin{equation*}        
    \max_{\norm{\ttf \delta}_2 \leq 1} e(\ttf{\delta},\tf{K}, \tf{\hat{K}})
        =
        \max_{\norm{\ttf \delta}_2 \leq 1}
        \ttf{\delta}^\top 
        \Big( \tf{\Phi}^\top \mathbf{C} \tf{\Phi} - \tf{\Hat{\Phi}}^\top \mathbf{C} \tf{\Hat{\Phi}} \Big) 
        \ttf{\delta}\,.
    \end{equation*}
By denoting $\tf{\Pi}(\tf\Phi):= \tf{\Phi}^\top \mathbf{C} \tf{\Phi} - \tf{\Hat{\Phi}}^\top \mathbf{C} \tf{\Hat{\Phi}} $ it holds that 
    \begin{equation} \label{eq:proof_dual_problem_spreg_lambda_max}
       \max_{\norm{\ttf \delta}_2 \leq 1} \ttf{\delta}^\top \tf{\Pi}(\tf\Phi) \ttf{\delta} = \lambda_{\max}(\tf{\Pi}(\tf\Phi))\,.
    \end{equation}
Problem~\eqref{eq:proof_dual_problem_spreg_lambda_max} can be reformulated in the following manner, as showcased in \cite[Sec. 2.2]{boyd1994linear}
      \[
        \min_\lambda \lambda \quad \st \quad  \lambda I - \tf{\Pi}(\tf{\Phi}) \succeq 0 \,.
    \]
    Finally, we retrieve \eqref{eq:regret_problem_dual_cost}, \eqref{eq:regret_problem_dual_pos_def_constraint} using the Schur complement of $(\lambda I  - \tf{\Pi}(\tf{\Phi}))$. Given the exactly reformulated cost, the minimal value of $\lambda$ in \eqref{eq:regret_problem_dual_cost} corresponds to the minimal value of $\spreg(\mathbf{K},\tf{\hat{K}})$ for all $\tf{K} \in \tf{\Theta}$. 3) If $\mathcal{S}$ is QI with respect to $\tf{G}$, then $\tf{\Theta} = \mathcal{S}$ by \eqref{eq:QI_equivalence_K_sparse}, and thus \eqref{eq:regret_problem_dual} is equivalent to \eqref{eq:minimize_spatial_regret_generic}.

\subsection{Implementation details}\label{ap:implementation_details}

The model of the system is described as follows. The state $x^i(\tau)$ of each mass $i$ is $
x^i(\tau) :=
\begin{bmatrix}
{p}^i(\tau) \\
{v}^i(\tau)
\end{bmatrix}$, where $p^i(\tau)$ and $v^i(\tau)$ represent the position and velocity of the mass $i \in \{1,\dots,10\}$ at time $\tau$, respectively. Each mass has a value of $m = \SI{1}{\kg}$ and is interconnected with its adjacent masses via dampers with $c = \SI{0.5}{\kg\per\second}$ and springs with $k = \SI{0.5}{\kg\per\second^2}$.
The dynamics of each mass $i$ is influenced by the states of its left and right neighbors $x^{i-1}(\tau)$ and $x^{i+1}(\tau)$. 
The end masses $x^1(\tau)$ and $x^{10}(\tau)$ do not present any left and right neighbors, respectively. 
The state evolution is described by 
\begin{equation*}
\scalemath{0.87}{
\dot{x}^i(\tau)
=
\begin{bmatrix}
0 & 0 & 0 & 1 & 0 & 0 \\
\frac{k}{m} & \frac{c}{m} & -\frac{2k}{m} & -\frac{2c}{m} & \frac{k}{m} & \frac{c}{m}
\end{bmatrix}
\begin{bmatrix} 
    x^{i-1}(\tau)\\
    x^i(\tau)\\
    x^{i+1}(\tau)
\end{bmatrix}
+
\begin{bmatrix}
0 \\
\frac{1}{m^i}
\end{bmatrix}u^i(\tau)\,.}
\end{equation*}
We discretize this system with a sampling time of $T_s = \SI{0.5}{\second}$, and set $\tf{C} = I$ to minimize input action and bring the states to zero.
The sparsity matrix $\tf{S}$ is defined as
\begin{equation*}
\tf{S} = \textnormal{Tril}(T) \otimes
\begin{bNiceArray}{>{\strut}llllllllll}[margin=1mm]
    \Block[borders={bottom,top,right,left,tikz={dashed,red}}]{1-3}{}1 & 1 & 1 & {} & {} & {} & {} & {} & \Block[borders={bottom,top,right,left,tikz={dashed,red}}]{5-2}{}1 & 1 \\
    {} & {} & \Block[borders={bottom,top,right,left,tikz={dashed,red}}]{1-3}{}1 & 1 & 1 & {} & {} & {} & 1 & 1 \\
    {} & {} & {} & {} & {} & \ddots & {} & {} & \vdots & \vdots \\
    {} & {} & {} & {} & {} & {} & \Block[borders={bottom,top,right,left,tikz={dashed,red}}]{1-3}{}1 & 1 & 1 & 1 \\
    {} & {} & {} & {} & {} & {} & {} & {} & \Block[borders={bottom,top,right,left,tikz={dashed,red}}]{1-2}{}1 & 1 \\
\end{bNiceArray}.
\end{equation*}
This means that the controller of each mass has access to the mass of its own agent plus the position of its right neighbor. Additionally, every controller has knowledge of the state $x^
{10}$ of the last agent.

\textbf{\textit{Reducing the Computation Load}}. Problem~\eqref{eq:regret_problem_dual} is a semidefinite optimization problem that exhibits a quadratic growth in the number of variables with respect to $T$: this can pose scalability challenges even for relatively short time windows. To mitigate this issue, inspired by infinite-horizon methods and following \cite{zhang2021stochastic}, we employ an approximation technique that favors scalability. Specifically, we introduce constraints on $\mathbf{\Phi}_u$ as described in \eqref{eq:achievability_constraint}. We impose a lower-block Toeplitz diagonal matrix structure, which we denote as~$\Bar{\mathbf{\Phi}}_u$
\begin{equation*}
\scalemath{0.9}{
\Bar{\tf{\Phi}}_u
:=
\begin{bmatrix}
\tf{\Phi}_{u,0} & \cdots & 0 \\
\vdots & \ddots &\vdots \\
\tf{\Phi}_{u,T-1} & \cdots& \tf{\Phi}_{u,0}
\end{bmatrix}\,.}
\end{equation*}
Consequently, $\Bar{\tf{\Phi}}_x$  is obtained by $\Bar{\mathbf{\Phi}}_x := (I + \tf{Z}\tf{A})^{-1}(I + \tf{Z}\tf{B}\Bar{\tf{\Phi}}_u)$ to guarantee~\eqref{eq:achievability_constraint}.
This approach significantly reduces the total number of variables in~\eqref{eq:regret_problem_dual} to $(mnT + 1)$, which now is linear in $T$.

} %

\addtolength{\textheight}{-12cm} 

\end{document}